\documentclass[a4paper,11pt]{article}
\usepackage{jheppub} 
\usepackage{lineno}
\usepackage{mathtools}
\usepackage{hyperref}
\usepackage{stmaryrd}
\usepackage{tikz-cd}
\usepackage{subcaption}
\usetikzlibrary{arrows.meta,positioning,calc}

\preprint{ZMP-HH/26-28, HU-EP-26/23-RTG}

\title{\boldmath Quantum Trigonometric Spin Ruijsenaars--Schneider Models from $K$-theoretic Coulomb Branches}

\author[a]{Gleb Arutyunov,}
\author[a]{Lukas Hardi,}
\author[b]{and Rob Klabbers.}
\affiliation[a]{II. Institut für Theoretische Physik, Universität Hamburg, \\ Luruper Chaussee 149, 22761 Hamburg, Germany}
\affiliation[b]{Humboldt-Universität zu Berlin, \\
Zum Großen Windkanal 2, 12489 Berlin, Germany}

\emailAdd{gleb.arutyunov@desy.de}
\emailAdd{lukas.hardi@desy.de}
\emailAdd{rob.klabbers@hu-berlin.de}

\abstract{We quantize the trigonometric spin Ruijsenaars--Schneider model of $N$ particles each with $\ell$ spin states using the recently developed description of the classical model in terms of the $K$-theoretic Coulomb branch of the 4d $\mathcal{N}=2$ quiver gauge theory for the necklace quiver with $\ell$ nodes of rank $N$. The main algebraic tool is an algebra of $L$-operators derived from abelianized monopole operators of minuscule charge, which turns the necklace quiver into an integrable spin chain by producing a family of commuting Hamiltonians. We show that the lowest Hamiltonian coincides with the first mode of the quantum determinant of the horizontal quantum loop algebra living inside the $K$-theoretic Coulomb branch algebra, whose Bethe subalgebra generates a maximal family of commuting Hamiltonians. Finally, we derive the commutation relations and quantum equations of motion of the quantized physical spin variables.}

\newcommand{\Z}{\mathbb{Z}}

\newcommand{\R}{\mathbb{R}}
\newcommand{\C}{\mathbb{C}}

\begin{document}
\maketitle
\flushbottom

\section{Introduction}

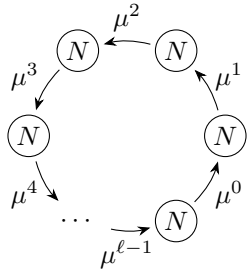
\begin{figure}[h]
    \centering
    \def\l{6}
    \def\radius{1.3cm}
    \begin{tikzpicture}[>=Stealth, every node/.style={font=\small}]
        \node[circle,draw,inner sep=1.5pt] (v1) at ({360*(0)/\l}:\radius) {$N$};
        \node[circle,draw,inner sep=1.5pt] (v2) at ({360*(1)/\l}:\radius) {$N$};
        \node[circle,draw,inner sep=1.5pt] (v3) at ({360*(2)/\l}:\radius) {$N$};
        \node[circle,draw,inner sep=1.5pt] (v4) at ({360*(3)/\l}:\radius) {$N$};
        \node (v5) at ({360*(4)/\l}:\radius) {$\cdots$};
        \node[circle,draw,inner sep=1.5pt] (v6) at ({360*(5)/\l}:\radius) {$N$};
        
        \foreach \i in {1,...,\l} {
            \pgfmathtruncatemacro{\next}{int(mod(\i,\l)+1)}
            \draw[->, shorten >=2pt, shorten <=2pt] (v\i) to[bend right=14] (v\next);
        }
        
        \node at ({(180+360*(0))/\l}:{1.2*\radius}) {\small $\mu^1$};
        \node at ({(180+360*(1))/\l}:{1.2*\radius}) {\small $\mu^2$};
        \node at ({(180+360*(2))/\l}:{1.2*\radius}) {\small $\mu^3$};
        \node at ({(180+360*(3))/\l}:{1.2*\radius}) {\small $\mu^4$};
        \node at ({(180+360*(4))/\l}:{1.2*\radius}) {\small $\mu^{\ell-1}$};
        \node at ({(180+360*(5))/\l}:{1.2*\radius}) {\small $\mu^0$};
    \end{tikzpicture}
    \caption{The necklace quiver with $\ell$ nodes indexed by $\alpha \in \Z/\ell\Z$, each of rank $N$. Every node contributes a $\mathrm{U}(N)$ factor to the gauge group and every arrow contributes a bifundamental hypermultiplet to the matter representation.}
    \label{fig:quiver}
\end{figure}

The trigonometric spin Ruijsenaars--Schneider (RS) model was introduced by 
Krichever and Zabrodin \cite{krichever:1995}. It is a  superintegrable system \cite{reshetikhin:2016} describing the dynamics of $N$ 
particles, each carrying $\ell$ 
internal
spin degrees of freedom, coupled via a coupling constant $t$. It was shown in \cite{arutyunov:2026} that the phase space of the model can be identified with the $K$-theoretic Coulomb branch \cite{braverman:2018} of the 4d $\mathcal{N}=2$ necklace quiver gauge theory on $\R^3 \times S^1$. The gauge group and matter representation of this theory are 
\begin{equation}
    G = \prod_{\alpha \in \Z/\ell\Z} \mathrm{U}(N)_\alpha, \qquad \mathbf{N} = \bigoplus_{\alpha \in \Z/\ell\Z} V_\alpha \otimes V_{\alpha+1}^\vee,
\end{equation}
where $V_\alpha$ denotes the defining representation of $\mathrm{U}(N)_\alpha$, and $V_\alpha^\vee$ its dual. Each bifundamental hypermultiplet $V_\alpha \otimes V_{\alpha+1}^\vee$ is assigned a mass deformation parameter $\mu^\alpha$, see figure \ref{fig:quiver}.
The coupling constant $t$ of the spin RS model is then identified with the product of the mass deformation parameters $\mu^\alpha$ for the bifundamental hypermultiplets, i.e.,
\begin{equation}
    t = \prod_{\alpha \in \Z/\ell\Z} \mu^\alpha.
\end{equation}

\begin{figure}[h]
    \centering
    \begin{tikzpicture}[>=Stealth,thick]
        \definecolor{defectblue}{RGB}{165,185,255}
        
        \draw[->] (0,-1.8) -- (0,2) node[above] {\small $\tau$};
        
        \draw[thick,->]
          (-0.23,0.15)
          arc[start angle=15+90,end angle=345+90,x radius=0.65,y radius=0.15];
        
        \foreach \yy in {-1.2,-0.4,0.4,1.2}
          {
          \draw[fill=defectblue] (0,\yy) circle (2pt);
          }
        
        \node at (0,-2.2) {$\mathbb{R}_\tau \times \C_z$};
        
        \node at (1.6,0.0) {\Large $\times_q$};
        
        \def\radius{0.9cm}
        \begin{scope}[shift={(3.5,0)}]
          \draw[white,fill=defectblue] (0,0) circle (1*\radius);
          \draw[fill=white] (0,0) circle (0.91*\radius);
        
          \node at (0,-2.2) {$S_\theta^1$};
        \end{scope}
        
        \tikzset{
          pics/torus/.style n args={3}{
            code = {
              \providecolor{pgffillcolor}{rgb}{1,1,1}
              \begin{scope}[
                  yscale=cos(####3),
                  outer torus/.style = {draw,line width/.expanded={\the\dimexpr2\pgflinewidth+####2*2},line join=round},
                  inner torus/.style = {draw=pgffillcolor,line width={####2*2}}
                ]
                \draw[outer torus] circle(####1);\draw[inner torus] circle(####1);
                \draw[outer torus] (180:####1) arc (180:360:####1);\draw[inner torus,line cap=round] (180:####1) arc (180:360:####1);
              \end{scope}
            }
          }
        }
        \end{tikzpicture}
    \caption{The 4d $\mathcal{N}=2$ necklace quiver gauge theory is placed on the twisted Cartesian product $\R_\tau \times \C_z \times_q S_\theta^1$, where the $\C_z$ plane is rotated by $q$ after winding once around the circle. The blue circles denote line defects whose fusion algebra is the $K$-theoretic Coulomb branch algebra. These defects are supported at points along the $\tau$-axis, located at the origin of the $\C_z$ plane, and wrap the circle $S_\theta^1$.}
    \label{fig:spacetime}
\end{figure}
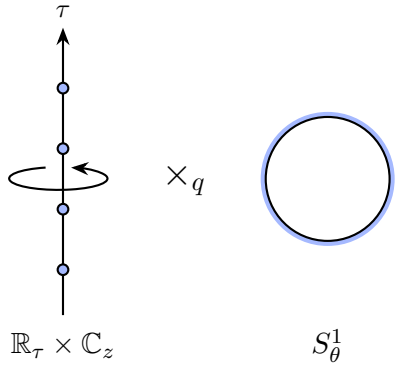

By the work of Braverman, Finkelberg, Nakajima, and others \cite{nakajima:2016,braverman:2018,braverman:2018b}, the algebra of functions on the $K$-theoretic Coulomb branch has a natural deformation quantization, depending on a parameter $q$, known as the quantized $K$-theoretic Coulomb branch algebra. This algebra admits a more explicit description in terms of a difference operator representation of quantum toroidal algebras \cite{tsymbaliuk:2023}. Physically, this deformation quantization corresponds to placing the 4d $\mathcal{N}=2$ theory on the $q$-twisted Cartesian product
\begin{equation}
    \R_\tau \times \C_z \times_q S_\theta^1 \coloneq \R_\tau \times (\C_z \times [0,2\pi]_\theta)/((z,0) \sim (qz,2\pi)).
\end{equation}
The quantized $K$-theoretic Coulomb branch algebra itself is the fusion algebra of line defects wrapping the circle $S_\theta^1$, as illustrated in figure \ref{fig:spacetime}. We describe its GKLO-type abelianization representation \cite{bullimore:2015}, \cite[remark 2.10]{tsymbaliuk:2023} via $q$-difference operators in section \ref{sec:CBAlg}.

In section \ref{sec:LOpAlgebra}, we quantize the classical $L$-operators $L^{\alpha\pm}$ from \cite{arutyunov:2026} associated with every arrow $\alpha \to \alpha+1$ in the necklace quiver. These $L$-operators are $N \times N$ matrices whose coefficients belong to a localization of the abelianized quantized $K$-theoretic Coulomb branch algebra and satisfy an RLL-type relation, see equations \eqref{eq:oneSiteLpAlg} and \eqref{eq:oneSiteLmAlg}. After conjugation with a diagonal matrix, our $L$-operators are related to the $L$-operators used in \cite{maruyoshi:2021}, where they were used to describe vacuum expectation values of Wilson--'t Hooft lines in the 4d $\mathcal{N}=2$ necklace quiver gauge theory.

Section \ref{sec:Hamiltonians} then uses the $L$-operator algebra to show the existence of a family of commuting Hamiltonians $H^\pm[n]$ coming from traces of powers of the monodromy around the necklace quiver, turning the necklace quiver into an integrable spin chain whose sites are the gauge nodes of the quiver. This family contains the Hamiltonian
\begin{equation}\label{eq:lowestHamiltonian}
    H^-[1] \propto \sum_{i_0,\dots,i_{\ell-1}=1}^N \Bigg( \prod_{\alpha=0}^{\ell-1} \prod_{k_\alpha (\neq i_\alpha)}^N \frac{Q_{i_{\alpha+1}}^{\alpha+1}-Q_{k_\alpha}^\alpha}{Q_{i_\alpha}^\alpha-Q_{k_\alpha}^\alpha} \Bigg) \prod_{\alpha=0}^{\ell-1} P_{i_\alpha}^\alpha,
\end{equation}
where $Q_i^\alpha$ are the vacuum expectation values of abelianized gauge multiplet scalars and $P_i^\alpha$ are $q$-shift operators, see section \ref{sec:CBAlg} for more detail. The Hamiltonian from equation \eqref{eq:lowestHamiltonian} is a natural generalization of the lowest Macdonald difference operator, which is the defining Hamiltonian of the spinless ($\ell=1$) trigonometric RS model. In terms of the 4d $\mathcal{N}=2$ necklace quiver gauge theory, the Hamiltonian $H^-[1]$ is the 't Hooft line which has charge $\square$ under all gauge nodes \cite{maruyoshi:2021}.

\smallskip
By the general theory of Braverman--Finkelberg--Nakajima, quantized homological Coulomb branch algebras are to be identified with truncated shifted Yangians \cite[appendix B]{braverman:2018b}. In the case of our necklace quiver, this statement reduces to an identification with the $N$-truncated (unshifted) affine Yangian $Y_\hbar^{(N)}(\dot{\mathfrak{gl}}_\ell)$ \cite{finkelberg:2014,nakajima:2017}. In line with this philosophy, the quantized $K$-theoretic Coulomb branch algebra for linear quivers is identified with truncated shifted quantum affine algebras \cite{finkelberg:2019}, 
while the corresponding algebra for necklace quivers should be identified with truncated shifted quantum toroidal algebras using the technology developed in \cite{tsymbaliuk:2023}. For our necklace quiver, the quantized $K$-theoretic Coulomb branch algebra should therefore be identified\footnote{In the spinless case $\ell=1$, the description of the quantized $K$-theoretic Coulomb branch algebra in terms of the quantum toroidal algebra was elucidated in \cite{zenkevich:2025}.} with the $N$-truncated (unshifted) quantum toroidal algebra $U_{q,t}^{(N)}(\ddot{\mathfrak{gl}}_\ell)$, see \cite{matsuo:2024} for a review. The quantum toroidal algebra $U_{q,t}(\ddot{\mathfrak{gl}}_\ell)$ (specialized to zero central charge) has a subalgebra isomorphic to the quantum loop algebra $U_q(\dot{\mathfrak{gl}}_\ell)$ called the \emph{horizontal subalgebra}. In section \ref{sec:quantumLoopAlg}, we exhibit Drinfeld--Jimbo and RTT generators of the horizontal quantum loop algebra and show that $H^-[1]$ may be identified with the first mode of its quantum determinant. The Hamiltonian $H^-[1]$ is thus part of the Bethe subalgebra of the quantum loop algebra, which is a maximal commutative subalgebra providing a large family of commuting Hamiltonians.

\smallskip
Lastly, section \ref{sec:spinVars} investigates the commutation relations of the physical spin variables of the trigonometric spin RS model and gives their quantum equations of motion. Originally, Krichever and Zabrodin \cite{krichever:1995} defined the trigonometric spin RS model via particle positions $Q_i^0$, particle momenta $P_i^0$ and spin covectors and vectors $a_i^\alpha,c_i^\alpha$ with $i=1,\dots,N$ and $\alpha \in \Z/\ell\Z$. Due to an extra symmetry in their equations of motion, the variables $a_i^0$ and $c_i^0$ can be expressed in terms of the other variables after symmetry reduction. The commutation relations of the unphysical spin variables $c^0$ are also considerably more complicated and are not quadratic like the commutation relations of the physical spin variables. This phenomenon was already observed at the classical level in \cite{arutyunov:2026} and in the rational case \cite{arutyunov:2025}. We give a summary of the physical variables in table \ref{table:physvars}. We find that the commutation relations of the physical spin variables are quadratic and are controlled by $R$-matrices. It is particularly notable that the physical spin vectors satisfy the simple exchange relation
\begin{equation}
    c_i^a c_i^b = R^{ab}(-1) c_i^b c_i^a, \qquad c_i^a c_j^b = R^{ab}(Q_j^0/Q_i^0) c_j^b c_i^a, \qquad (i \neq j)
\end{equation}
where $R(u)$ is the standard trigonometric $R$-matrix for $\mathfrak{gl}_{\ell-1}$, see equation \eqref{eq:trigRMat}.
\begin{table}[h]
\centering
\begin{tabular}{l|l}
    \hline
    physical variable & physical meaning \\
    \hline
    $Q_i^0$ & position of the $i$th particle \\
    $P_i^0$ & momentum of the $i$th particle \\
    $a_i \coloneq (a_i^1,\dots,a_i^{\ell-1})$ & physical spin covector of the $i$th particle \\
    $c_i \coloneq (c_i^1,\cdots,c_i^{\ell-1})^t$ & physical spin vector of the $i$th particle \\
    \hline
\end{tabular}
\caption{List of the $2N\ell$ physical variables of the trigonometric spin RS model.\label{table:physvars}}
\end{table}

\paragraph{Related work.} The classical trigonometric spin RS model was also constructed via the method of quasi-Hamiltonian reduction \cite{chalykh:2020} as a multiplicative quiver variety and Poisson reduction \cite{fairon:2021,arutyunov:2019} from the Heisenberg double. This construction can be thought of as the mirror dual of the construction of the classical trigonometric spin RS model from $K$-theoretic Coulomb branches put forth in \cite{arutyunov:2026}. The quasi-Hamiltonian reduction approach was generalized to the case of multiple sets of spin variables in \cite{fairon:2025}. Following the construction of commuting Hamiltonians in \cite{bernard:1993,cherednik:1994}, another a priori distinct but seemingly closely related
quantum trigonometric spin RS model was defined in \cite{uglov:1995} and significantly simplified in
\cite{lamers:2022}. Namely, it also arises as a representation of the quantum toroidal algebra $U_{q,t}(\ddot{\mathfrak{gl}}_\ell)$, but on an $N$-fold tensor product of vector representations, giving rise to the $q$-deformed Haldane--Shastry spin chain \cite{bernard:1993,uglov:1995,lamers:2022} after freezing. This line of research also allows for the definition of a quantum elliptic spin RS model \cite{klabbers:2024b}, which upon freezing gives rise to the $q$-deformed Inozemtsev spin chain \cite{inozemtsev:1990,klabbers:2024b}. The Inozemtsev spin chain regained interest due to its appearance in integrability of $\mathcal{N}=4$ super Yang--Mills theory \cite{serban:2004,serban:2011}.

\paragraph{Notation.} We use Latin lower indices $i,j,\dots$ ranging from $1$ to $N$ to denote particle/color indices and Greek upper indices $\alpha,\beta,\dots$ ranging over $\Z/\ell\Z$ to denote spin/gauge node indices. We also use Latin letters $a,b,\dots$ to denote auxiliary space indices. We let $e_{ij}$ denote $N \times N$ matrix units and $e^{\alpha\beta}$ denote $\ell \times \ell$ matrix units. Parentheses in summation
or product ranges indicate the enclosed index is not summed over, e.g. $\sum_{i(\neq j)}^N = \sum_{\substack{i=1 \\ i \neq j}}^N$.

\section{The $K$-theoretic Coulomb branch algebra}\label{sec:CBAlg}

In this section, we construct generators of the abelianized quantized $K$-theoretic Coulomb branch algebra of the necklace quiver following \cite[remark 2.10]{tsymbaliuk:2023}. The generators can be expressed in terms of the exponentiated vacuum expectation values of abelianized gauge multiplet scalars $Q_i^\alpha$, where $\alpha \in \Z/\ell\Z$ labels a node of the quiver and $i = 1,\dots,N$ is a color index for the corresponding gauge group factor $U(N)_\alpha$, as well as the $q$-difference operators $P_i^\alpha$, which satisfy the commutation relations
\begin{equation}
    Q_i^\alpha Q_j^\beta = Q_j^\beta Q_i^\alpha, \qquad P_i^\alpha Q_j^\beta = q^{\delta_{ij} \delta^{\alpha\beta}} Q_j^\beta P_i^\alpha, \qquad P_i^\alpha P_j^\beta = P_j^\beta P_i^\alpha.
\end{equation}
The abelianized monopole operators of minuscule coweight/magnetic charge, which together with the exponentiated vacuum expectation values $Q_i^\alpha$ generate the abelianized quantized $K$-theoretic Coulomb branch algebra, can then be expressed in terms of the $q$-difference operators as
\begin{align}
    u_i^{\alpha+} &\coloneq \prod_{k=1}^N (Q_k^\alpha/Q_k^{\alpha+1})^{1/2} \frac{\prod_{k=1}^N (1-(\mu^\alpha)^{-1} Q_k^{\alpha+1}/Q_i^\alpha)}{\prod_{k(\neq i)} (1-Q_k^\alpha/Q_i^\alpha)} P_i^\alpha, \\
    u_i^{\alpha-} &\coloneq \prod_{k=1}^N (Q_k^\alpha/Q_k^{\alpha-1})^{1/2} \frac{\prod_{k=1}^N (1-\mu^{\alpha-1} Q_k^{\alpha-1}/Q_i^\alpha)}{\prod_{k(\neq i)} (1-Q_k^\alpha/Q_i^\alpha)} (P_i^\alpha)^{-1}.
\end{align}
We now describe their commutation relations. To this end, we define the structure function
\begin{equation}
    g_\kappa^\pm(z) \coloneq \frac{q^{\mp\kappa/2} z - 1}{z - q^{\mp\kappa/2}}.
\end{equation}
Given $(i,\alpha) \neq (j,\beta)$, we then find
\begin{align}
    u_i^{\alpha\pm} u_j^{\beta\pm} &= g_{\kappa^{\alpha\beta}}^\pm\big((q^{-1/2} \mu^\alpha)^{\delta^{\alpha+1,\beta}} (q^{-1/2} \mu^\beta)^{-\delta^{\alpha,\beta+1}} Q_{ij}^{\alpha\beta}\big) u_j^{\beta\pm} u_i^{\alpha\pm}, & (\ell > 2) \\
    u_i^{\alpha\pm} u_j^{\beta\pm} &= g_{2}^\pm\big(Q_{ij}^{\alpha\beta}\big) u_j^{\beta\pm} u_i^{\alpha\pm}, & (\alpha = \beta,\ell = 2) \\
    u_i^{\alpha\pm} u_j^{\beta\pm} &= g_{-1}^\pm\big(q^{-1/2}\mu^\alpha Q_{ij}^{\alpha\beta}\big) g_{-1}^\pm\big(q^{1/2}(\mu^\beta)^{-1} Q_{ij}^{\alpha\beta}\big) u_j^{\beta\pm} u_i^{\alpha\pm}, & (\alpha \neq \beta,\ell = 2) \\
    u_i^{0\pm} u_j^{0\pm} &= g_{2}^\pm\big(Q_{ij}^{00}\big) g_{-1}^\pm\big(q^{-1/2}\mu^0 Q_{ij}^{00}\big) g_{-1}^\pm\big(q^{1/2}(\mu^0)^{-1} Q_{ij}^{00}\big) u_j^{0\pm} u_i^{0\pm}, & (\ell = 1)
\end{align}
where we have used the abbreviation $Q_{ij}^{\alpha\beta} \coloneq Q_i^\alpha/Q_j^\beta$ and $\kappa^{\alpha\beta} \coloneq 2\delta^{\alpha\beta} - \delta^{\alpha+1,\beta} - \delta^{\alpha,\beta+1}$ is the Cartan matrix of the necklace quiver.

\section{The $L$-operator algebra}\label{sec:LOpAlgebra}

\begin{figure}[t]
    \centering
    \begin{subfigure}{0.32\textwidth}
    \centering
    \begin{tikzpicture}[>=Stealth,thick,scale=0.75]
        \draw[->] (0.3,-1) -- (0.3,1);
        \draw[->] (-1,-1) -- (1,1);
        \draw[->] (1,-1) -- (-1,1);
        \node[below] at (-1,-1) {\footnotesize $\alpha$};
        \node[below] at (0.3,-1) {\footnotesize $\beta$};
        \node[below] at (1,-1) {\footnotesize $\gamma$};
        
        \node at (1.5,0) {$=$};
        
        \begin{scope}[xshift=3cm]
            \draw[->] (-0.3,-1) -- (-0.3,1);
            \draw[->] (-1,-1) -- (1,1);
            \draw[->] (1,-1) -- (-1,1);
            \node[below] at (-1,-1) {\footnotesize $\alpha$};
            \node[below] at (-0.3,-1) {\footnotesize $\beta$};
            \node[below] at (1,-1) {\footnotesize $\gamma$};
        \end{scope}
    \end{tikzpicture}
    \caption{}
    \end{subfigure}
    \begin{subfigure}{0.32\textwidth}
    \centering
    \begin{tikzpicture}[>=Stealth,thick,scale=0.75]
        \draw[gray!30!white,fill=gray!30!white] (-1,-1) rectangle (1,1);
        \draw[gray!30!white,fill=gray!30!white] (2,-1) rectangle (4,1);
    
        \draw[->] (0.3,-1) -- (0.3,1);
        \draw[->] (-1,-1) -- (1,1);
        \draw[->] (1,-1) -- (-1,1);
        \node[below] at (-1,-1) {\footnotesize $\alpha$};
        \node[below] at (0.3,-1) {\footnotesize $\beta$};
        \node[below] at (1,-1) {\footnotesize $\gamma$};
        \node[gray] at (0.6,0) {\scriptsize $Q$};
        
        \node at (1.5,0) {$=$};
        
        \begin{scope}[xshift=3cm]
            \draw[->] (-0.3,-1) -- (-0.3,1);
            \draw[->] (-1,-1) -- (1,1);
            \draw[->] (1,-1) -- (-1,1);
            \node[below] at (-1,-1) {\footnotesize $\alpha$};
            \node[below] at (-0.3,-1) {\footnotesize $\beta$};
            \node[below] at (1,-1) {\footnotesize $\gamma$};
            \node[gray] at (0.5,0) {\scriptsize $Q$};
        \end{scope}
    \end{tikzpicture}
    \caption{}
    \end{subfigure}
    \begin{subfigure}{0.32\textwidth}
    \centering
    \begin{tikzpicture}[>=Stealth,thick,scale=0.75]
        \draw[gray!30!white,fill=gray!30!white] (-1,-1) rectangle (1,0);
        \draw[gray!30!white,fill=gray!30!white] (2,-1) rectangle (4,0);
    
        \draw[->] (-0.7,-1) -- (1,0.7);
        \draw[->] (0.7,-1) -- (-1,0.7);
        \node[below] at (-0.7,-1) {\footnotesize $\alpha$};
        \node[below] at (0.7,-1) {\footnotesize $\beta$};
        \node[gray] at (0.5,-0.3) {\scriptsize $Q$};
        
        \node at (1.5,0) {$=$};

        \begin{scope}[xshift=3cm]
            \draw[->] (-1,-0.7) -- (0.7,1);
            \draw[->] (1,-0.7) -- (-0.7,1);
            \node[left] at (-1,-0.7) {\footnotesize $\alpha$};
            \node[right] at (1,-0.7) {\footnotesize $\beta$};
            \node[gray] at (0.82,-0.225) {\scriptsize $Q$};
        \end{scope}
    \end{tikzpicture}
    \caption{}
    \end{subfigure}
    \caption{Diagrammatic representation of the three relations satisfied by the $R$-matrices. (a) The Yang--Baxter equation for the $R$-matrix $R^{\alpha\beta}(q)$, which is depicted as a crossing of two lines labeled by $\alpha$ and $\beta$ on a white background. (b) The dynamical Yang--Baxter equation for the dynamical $R$-matrix $\underline R^{\alpha\beta}(q)$, which is depicted as a crossing of two lines labeled by $\alpha$ and $\beta$ on a gray background. The faces with a gray background are labeled by the dynamical parameters $Q$, with their values determined from the reference face (labeled $Q$) by $q$-shifting $Q^\alpha$ when passing over a line labeled $\alpha$. (c) The twist relation for the dynamical twist $\bar R^{\alpha\beta}(q)$, which is depicted as two lines labeled by $\alpha$ and $\beta$ crossing from a gray to a white background. Composition is read from bottom to top.}
    \label{fig:Rmatrelations}
\end{figure}
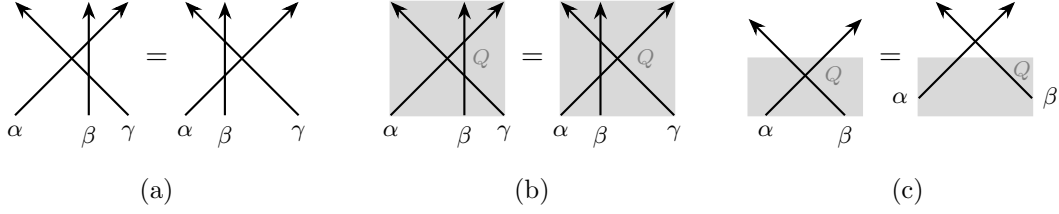

The main tool for obtaining the classical trigonometric spin RS model from the $K$-theoretic Coulomb branch algebra is the $L$-operator algebra introduced in \cite{arutyunov:2026}.
In this section 
we quantize this algebra. To every arrow $\alpha \to \alpha+1$ in the necklace quiver, we associate two $L$-operators $L^{\alpha\pm}$, which are $N \times N$ matrices with coefficients
\begin{align}
    L_{ij}^{\alpha+} &\coloneq \frac{q^{1/2}}{1-(\mu^\alpha)^{-1} Q_i^{\alpha+1}/Q_j^\alpha} u_j^{\alpha+}, \label{eq:LpOp} \\
    L_{ij}^{\alpha-} &\coloneq \frac{q^{-1/2}}{\mu^\alpha Q_i^\alpha/Q_j^{\alpha+1}-1} u_j^{\alpha+1,-}. \label{eq:LmOp}
\end{align}
In \cite{arutyunov:2026} it was shown that 
the Poisson brackets of the classical counterparts of these $L$-operators are quadratic in $L^{\alpha\pm}$ and can be written in terms of classical $r$-matrices. Assuming the operator ordering prescribed by equations \eqref{eq:LpOp} and \eqref{eq:LmOp}, we obtain the following commutation relations
\begin{align}
    R_{ab}^{\alpha+1,\beta+1}(q^{-1}) L_b^{\beta+} \bar R_{ab}^{\alpha+1,\beta}(q^{-1})^{-1} L_a^{\alpha+} &= L_a^{\alpha+} \bar R_{ba}^{\beta+1,\alpha}(q^{-1})^{-1} L_b^{\beta+} \underline R_{ab}^{\alpha\beta}(q^{-1}), \label{eq:oneSiteLpAlg} \\
    R_{ab}^{\alpha\beta}(q) L_b^{\beta-} \bar R_{ab}^{\beta+1,\alpha}(q)^{-1} L_a^{\alpha-} &= L_a^{\alpha-} \bar R_{ba}^{\beta,\alpha+1}(q)^{-1} L_b^{\beta-} \underline R_{ab}^{\alpha+1,\beta+1}(q). \label{eq:oneSiteLmAlg}
\end{align}
These relations provide a quantization of the aforementioned Poisson brackets.
Here, $a$ and $b$ denote auxiliary spaces and we introduce the following quantum $R$-matrices:
\begin{align}\nonumber
    R^{\alpha\beta}(q) \coloneq{}& 1 + \delta^{\alpha\beta} (1-q^{-1}) \times \\ & \Bigg[ \sum_{i \neq j} \bigg( \frac{1}{Q_i^\alpha/Q_j^\alpha-1} e_{ii} - \frac{1}{1-Q_j^\alpha/Q_i^\alpha} e_{ij} \bigg) \otimes (e_{jj}-e_{ji}) + X - X_{21} \Bigg], \\
    \bar R^{\alpha\beta}(q) \coloneq{}& 1 + \delta^{\alpha\beta} \bigg[ (q^{-1/2}-1) + \sum_{i \neq j} \frac{q^{1/2}-q^{-1/2}}{1 - q Q_j^\alpha/Q_i^\alpha} (e_{ii}-e_{ij}) \otimes e_{jj} + (q^{1/2}-q^{-1/2}) X \bigg], \\
    \underline R^{\alpha\beta}(q) \coloneq{}& 1 + \delta^{\alpha\beta} (1-q^{-1}) \sum_{i \neq j} \frac{1}{1-Q_j^\alpha/Q_i^\alpha} (e_{ij} \otimes e_{ji} - e_{ii} \otimes e_{jj}),
\end{align}
with $X = \sum_{i,j=1}^N e_{ij} \otimes e_{jj}$. These $R$-matrices are identical to the evaluation at zero of the spectral parameter dependent 
$R$-matrices found in \cite{arutyunov:2019b} for the spinless case up to a prefactor for $\bar R^{\alpha\beta}(q)$. It should be noted that $R^{\alpha\beta}(q)$ satisfies the Yang--Baxter equation
\begin{equation}
    R_{ab}^{\alpha\beta}(q) R_{ac}^{\alpha\gamma}(q) R_{bc}^{\beta\gamma}(q) = R_{bc}^{\beta\gamma}(q) R_{ac}^{\alpha\gamma}(q) R_{ab}^{\alpha\beta}(q),
\end{equation}
and $\underline R^{\alpha\beta}(q)$ satisfies the dynamical Yang--Baxter equation \begin{equation}
    \underline R_{ab}^{\alpha\beta}(q) (P_b^\beta)^{-1} \underline R_{ac}^{\alpha\gamma}(q) P_b^\beta \underline R_{bc}^{\beta\gamma}(q) = (P_a^\alpha)^{-1} \underline R_{bc}^{\beta\gamma}(q) P_a^\alpha \underline R_{ac}^{\alpha\gamma}(q) (P_c^\gamma)^{-1} \underline R_{ab}^{\alpha\beta}(q) P_c^\gamma,
\end{equation}
where the role of the dynamical parameters is played by the variables $Q_i^\alpha$. Finally, $\bar R^{\alpha\beta}(q)$ is the corresponding dynamical twist, i.e.
\begin{equation}\label{eq:dynamicalTwistRel}
    \bar R^{\alpha\beta}(q) \underline R^{\alpha\beta}(q) = R^{\alpha\beta}(q) \bar R_{21}^{\alpha\beta}(q).
\end{equation}
These relations are shown diagrammatically in figure \ref{fig:Rmatrelations}.

If we define the total $L$-operators to be
\begin{align}
    L^{\text{tot}+} \coloneq L^{\ell-1,+} \cdots L^{0,+}, \qquad L^{\text{tot}-} \coloneq L^{0-} \cdots L^{\ell-1,-},
\end{align}
the relations \eqref{eq:oneSiteLpAlg} and \eqref{eq:oneSiteLmAlg} together with equation \eqref{eq:dynamicalTwistRel} imply
\begin{equation}\label{eq:totalAlg}
    R_{ab}^{00}(q^{\mp}) L_b^{\text{tot}\pm} \bar R_{ab}^{00}(q^{\mp})^{-1} L_a^{\text{tot}\pm} = L_a^{\text{tot}\pm} \bar R_{ba}^{00}(q^{\mp})^{-1} L_b^{\text{tot}\pm} \underline R_{ab}^{00}(q^{\mp}),
\end{equation}
which exactly reproduces the $L$-operator algebra of the spinless model found in \cite{arutyunov:2019b}.

\section{Commuting Hamiltonians}\label{sec:Hamiltonians}

The commutation relations \eqref{eq:totalAlg} of the total $L$-operator coincide with those obtained in \cite{arutyunov:2019b} for the $L$-operator of the spinless model. It follows that a family of commuting Hamiltonians is given by the same trace formula as in \emph{loc.\ cit.}:
\begin{equation}\label{eq:traceFormula}
    H^\pm[n] \coloneq \operatorname{Tr}_{ab} P_{ab}^{t_b} L_a^{\text{tot}\pm} (\bar R_{ba}^{00}(q^{\mp})^{t_b} R_{ab}^{00}(q^{\mp})^{t_b} L_a^{\text{tot}\pm})^{n-1},
\end{equation}
where $P = \sum_{i,j=1}^N e_{ij} \otimes e_{ji}$ and the superscript $t_b$ denotes transposition in the auxiliary space $b$. In particular, we have $H^-[1] = \operatorname{Tr} L^{\text{tot}-}$, which coincides with the defining Hamiltonian of the trigonometric spin RS model introduced in \cite{arutyunov:2026} and generates at the classical level the equations of motion found by Krichever and Zabrodin \cite{krichever:1995}. We remark that in terms of the 4d $\mathcal{N}=2$ necklace quiver gauge theory, the Hamiltonian $H^-[1]$ corresponds to the 't Hooft line which has charge $\square$ under all gauge nodes \cite{maruyoshi:2021}.

\smallskip
The Hamiltonians \eqref{eq:traceFormula} can be rewritten in terms of the generalized Macdonald operators
\begin{equation}\label{eq:macdonald}
    \begin{aligned}
    S^+[n] &\coloneq t^{-n(n-1)/2} \!\!\!\!\!\!\!\!\!\! \sum_{\substack{I_0,\dots,I_{\ell-1} \subseteq \{ 1,\dots,N \} \\ |I_\alpha|=n}} \Bigg( \prod_{\alpha=0}^{\ell-1} \frac{\prod_{i \in I_{\alpha-1},k \notin I_\alpha} (Q_i^{\alpha-1}-(\mu^{\alpha-1})^{-1} Q_k^\alpha)}{\prod_{i \in I_{\alpha},k \notin I_\alpha} (Q_i^\alpha-Q_k^\alpha)} \Bigg) \prod_{\alpha=0}^{\ell-1} \prod_{i \in I_\alpha} P_i^\alpha, \\
    S^-[n] &\coloneq t^{n(n-1)/2} \!\!\!\!\!\! \sum_{\substack{I_0,\dots,I_{\ell-1} \subseteq \{ 1,\dots,N \} \\ |I_\alpha|=n}} \Bigg( \prod_{\alpha=0}^{\ell-1} \frac{\prod_{i \in I_{\alpha+1},k \notin I_\alpha} (Q_i^{\alpha+1}-\mu^\alpha Q_k^\alpha)}{\prod_{i \in I_{\alpha},k \notin I_\alpha} (Q_i^\alpha-Q_k^\alpha)} \Bigg) \prod_{\alpha=0}^{\ell-1} \prod_{i \in I_\alpha} (P_i^\alpha)^{-1}.
    \end{aligned}
\end{equation}
Specifically, we have the identification
\begin{equation}\label{eq:determinant}
    H^\pm[n] = q^{\pm 1/2}
    \begin{vmatrix}
        S^\pm[1] & 1 & 0 & \cdots & 0 \\
        [2]_{q^{\pm 1}} S^\pm[2] & S^\pm[1] & 1 & \cdots & 0 \\
        \vdots & \vdots & \vdots & \ddots & \vdots \\
        [n]_{q^{\pm 1}} S^\pm[n] ~&~ S^\pm[n-1] ~&~ S^\pm[n-2] & \cdots & S^\pm[1]
    \end{vmatrix},
\end{equation}
where we make use of the $q$-numbers $[n]_q = \sum_{k=0}^{n-1} q^k$. This suggests that $S^\pm[n]$ are the elementary symmetric counterparts of the power sums $H^\pm[n]$. Indeed, we find
\begin{equation}\label{eq:spectralCurve}
    : {\det(q^{\mp\ell/2} L^{\text{tot}\pm}+z)} : \, = \sum_{n=0}^N z^{N-n} S^\pm[n],
\end{equation}
where $::$ means that the $P_i^\alpha$ are normal ordered to the right of the $Q_i^\alpha$. Note that introducing a spectral parameter dependence in $L^{\text{tot}\pm}$ would turn equation \eqref{eq:spectralCurve} into the quantum spectral curve of the trigonometric spin RS model.

To finish the discussion of the Hamiltonians, let us note that we can build generators of the 1+1d Poincaré Lie algebra from the lowest Hamiltonians $H^\pm[1]$ as follows:
\begin{equation}
    \mathbf{H} \coloneq H^+[1] + H^-[1], \qquad \mathbf{P} \coloneq H^+[1] - H^-[1], \qquad \mathbf{B} \coloneq \sum_{i=1}^N \log Q_i^0.
\end{equation}
They obey the expected relations
\begin{equation}
    [\mathbf{H},\mathbf{P}] = 0, \qquad [\mathbf{H},\mathbf{B}] = \log(q) \mathbf{P}, \qquad [\mathbf{P},\mathbf{B}] = \log(q) \mathbf{H},
\end{equation}
and the boost generator $\mathbf{B}$ more generally satisfies
\begin{equation}
    [H^\pm[n],\mathbf{B}] = \pm n \log(q) H^\pm[n], \qquad [S^\pm[n],\mathbf{B}] = \pm n \log(q) S^\pm[n].
\end{equation}
This repeats the known pattern of the spinless trigonometric RS model.

\section{Quantum loop algebra}\label{sec:quantumLoopAlg}

In this section, we exhibit a quantum loop algebra structure inside the $K$-theoretic Coulomb branch algebra. As discussed in the introduction, the general theory of \cite[appendix B]{braverman:2018b} and the technology developed in \cite{tsymbaliuk:2023} suggest that the $K$-theoretic Coulomb branch of the necklace quiver should be identified with the $N$-truncated quantum toroidal algebra $U_{q,t}^{(N)}(\ddot{\mathfrak{gl}}_\ell)$. Specializing to zero central charge, the quantum toroidal algebra has two subalgebras isomorphic to the $N$-truncated quantum loop algebra $U_q^{(N)}(\dot{\mathfrak{gl}}_\ell)$, called \emph{horizontal} and \emph{vertical}. Here, we exhibit the horizontal quantum loop subalgebra.

\smallskip
The quantum loop algebra $U_q(\dot{\mathfrak{gl}}_\ell)$ has various useful presentations. Firstly, it may be presented in the Drinfeld--Jimbo generators
\begin{equation}
    E^\alpha, \qquad (t^\alpha)^{\pm 1}, \qquad F^\alpha,
\end{equation}
which satisfy relations analogous to the relations for $U_q(\mathfrak{gl}_\ell)$ \cite{finkelberg:2019}:
\begin{align}
    (t^\alpha)^{\pm 1} (t^\alpha)^{\mp 1} &= 1, \label{eq:DJrel1} \\[5pt]
    [t^\alpha,t^\beta] &= 0, \\
    [E^\alpha,F^\beta] &= \delta^{\alpha\beta} \frac{K^\alpha-(K^\alpha)^{-1}}{q^{1/2}-q^{-1/2}}, \\
    E^\alpha t^\beta &= q^{(-\delta^{\alpha\beta}+\delta^{\alpha+1,\beta})/2} t^\beta E^\alpha \\[5pt]
    F^\alpha t^\beta &= q^{(\delta^{\alpha\beta}-\delta^{\alpha+1,\beta})/2} t^\beta F^\alpha, \\[5pt]
    E^\alpha E^\beta &= E^\beta E^\alpha, \qquad \alpha-\beta \neq 0,\pm1, \\[5pt]
    F^\alpha F^\beta &= F^\beta F^\alpha, \qquad \alpha-\beta \neq 0,\pm1,
\end{align}
where $K^\alpha \coloneq t^\alpha (t^{\alpha-1})^{-1}$, as well as the $q$-Serre relations
\begin{align}
    (E^\alpha)^2 E^{\alpha\pm1} + E^{\alpha\pm1} (E^\alpha)^2 &= (q^{1/2}+q^{-1/2}) E^\alpha E^{\alpha\pm1} E^\alpha, \\[5pt]
    (F^\alpha)^2 F^{\alpha\pm1} + F^{\alpha\pm1} (F^\alpha)^2 &= (q^{1/2}+q^{-1/2}) F^\alpha F^{\alpha\pm1} F^\alpha,
\end{align}
in the case $\ell > 2$, and
\begin{align}
    (E^\alpha)^3 E^{\alpha\pm1} - E^{\alpha\pm1} (E^\alpha)^3 &= (q^{-1} + 1 + q) \Big[ (E^\alpha)^2 E^{\alpha\pm1} E^\alpha - E^\alpha E^{\alpha\pm1} (E^\alpha)^2 \Big], \\[5pt]
    (F^\alpha)^3 F^{\alpha\pm1} - F^{\alpha\pm1} (F^\alpha)^3 &= (q^{-1} + 1 + q) \Big[ (F^\alpha)^2 F^{\alpha\pm1} F^\alpha - F^\alpha F^{\alpha\pm1} (F^\alpha)^2 \Big], \label{eq:DJrel2}
\end{align}
in the case $\ell = 2$. The case $\ell=1$ does not have Serre relations.

\smallskip
Secondly, the RTT presentation \cite{molev} of the quantum loop algebra $U_q(\dot{\mathfrak{gl}}_\ell)$ is given in terms of generator matrices
\begin{equation}
    T^\pm(u) = \sum_{\alpha,\beta=1}^\ell \sum_{n \geq 0} T^\pm[n]^{\alpha\beta} \otimes u^{\mp n} e^{\alpha\beta} \in U_q(\dot{\mathfrak{gl}}_\ell) \otimes \operatorname{End}(\C^\ell)\llbracket u^{\mp 1} \rrbracket,
\end{equation}
which satisfy the additional requirements 
\begin{equation}
    T^\pm[0]^{\alpha\alpha} T^\mp[0]^{\alpha\alpha} = 1, \qquad T^\pm[0]^{\alpha\beta} = 0, \qquad \alpha \gtrless \beta,
\end{equation}
as well as the RTT relations
\begin{align}\label{eq:RTT}
    R^{ab}(u/v) T^\pm(u)^a T^\pm(v)^b &= T^\pm(v)^b T^\pm(u)^a R^{ab}(u/v), \\
    R^{ab}(u/v) T^-(u)^a T^+(v)^b &= T^+(v)^b T^-(u)^a R^{ab}(u/v),
\end{align}
where we have used the standard unitary trigonometric $R$-matrix for $\mathfrak{gl}_\ell$:
\begin{equation}\label{eq:trigRMat}
    \begin{aligned}
        R(u) ={}& \frac{q^{1/2}-q^{-1/2} u}{q^{-1/2}-q^{1/2} u} \sum_\alpha e^{\alpha\alpha} \otimes e^{\alpha\alpha} + \frac{1-u}{q^{-1/2}-q^{1/2} u} \sum_{\alpha\neq\beta} e^{\alpha\alpha} \otimes e^{\beta\beta} \\
        &+ \frac{q^{1/2}-q^{-1/2}}{q^{-1/2}-q^{1/2} u} \sum_{\alpha>\beta} e^{\alpha\beta} \otimes e^{\beta\alpha} + \frac{q^{1/2} u-q^{-1/2} u}{q^{-1/2}-q^{1/2} u} \sum_{\alpha<\beta} e^{\alpha\beta} \otimes e^{\beta\alpha}.
    \end{aligned}
\end{equation}
The isomorphism between the Drinfeld--Jimbo and the RTT presentation is explicitly given in \cite{finkelberg:2019} for the case of $U_q(\mathfrak{gl}_\ell)$ and has a natural generalization for the case of the quantum loop algebra $U_q(\dot{\mathfrak{gl}}_\ell)$.

\smallskip
We now build a representation of the quantum loop algebra $U_q(\dot{\mathfrak{gl}}_\ell)$ from the abelianized monopole operators $u_i^\alpha$. Indeed, it may be checked that the generators
\begin{align}
    t^\alpha &\coloneq \prod_{k=1}^N ((\mu^{\alpha-1})^{-1} Q_k^\alpha/Q_k^{\alpha-1})^{1/2}, \label{eq:DJgen1} \\
    E^{\alpha} &\coloneq \frac{(\mu^{\alpha-1})^{-N/2}}{q^{-1/2}-q^{1/2}} \sum_{i=1}^N u_i^{\alpha+}, \\
    F^{\alpha} &\coloneq \frac{(\mu^\alpha)^{N/2}}{q^{1/2}-q^{-1/2}} \sum_{i=1}^N u_i^{\alpha-}, \label{eq:DJgen2}
\end{align}
satisfy the relations \eqref{eq:DJrel1}--\eqref{eq:DJrel2}. More generally, we can use the $L$-operator algebra to define the combinations
\begin{align}
    E^{\alpha,\beta} \coloneq (\mu^{\alpha-1})^{-1} u^{\beta+} L^{\beta-1,+} \cdots L^{\alpha+} e, \qquad F^{\alpha,\beta} \coloneq \mu^\beta u^{\alpha-} L^{\alpha-} \cdots L^{\beta-1,-} e,
\end{align}
where we use the covector $u^{\alpha\pm} \coloneq (u_1^{\alpha\pm},\dots,u_N^{\alpha\pm})$ and the vector $e = (1,\dots,1)^t$. In terms of the 4d $\mathcal{N}=2$ necklace quiver gauge theory, these may be thought of as monopole operators with fundamental charge under the gauge nodes $\alpha$ to $\beta$, see \cite[\S 6.6.1]{bullimore:2015}. From the integrable spin chain perspective, they may be thought of as the propagation of an excitation along the spin chain from site $\alpha$ to site $\beta$. Using the isomorphism between the Drinfeld--Jimbo and the RTT presentation \cite{finkelberg:2019}, we then find the zeroth and first modes of the generator matrices $T^\pm(u)$ to be of the form
\begin{equation}
    T^+[0]^{\alpha\beta} =
    \begin{cases}
        (t^\alpha)^{-1} E^{\alpha,\beta-1}, & \alpha < \beta,  \\
        (t^\alpha)^{-1}, & \alpha = \beta, \\
        0, & \alpha > \beta,
    \end{cases} \qquad \qquad 
    T^-[0]^{\alpha\beta} =
    \begin{cases}
        F^{\beta,\alpha-1} t^\beta, & \alpha > \beta, \\
        t^\beta, & \alpha = \beta, \\
        0, & \alpha < \beta,
    \end{cases}
\end{equation}
and
\begin{align}
    T^+[1]^{\alpha\beta} &= q^{-(1-\delta^{\alpha\beta})/2} (t^\alpha)^{-1} E^{\alpha,\beta+\ell-1}, \qquad & T^-[1]^{\alpha\beta} &= q^{(1-\delta^{\alpha\beta})/2} F^{\beta,\alpha+\ell-1} t^\beta.
\end{align}
One may check that the RTT relations \eqref{eq:RTT} of the quantum loop algebra are indeed satisfied to first order in $u$ and $v$. Now, we can use the quantum Leibniz formula
\begin{equation}
    \operatorname{qdet}^\pm(u) = \sum_{n=0}^\infty \operatorname{qdet}^\pm[n] u^{\mp n} = \sum_{\sigma \in S_\ell} (-q^{1/2})^{\ell(\sigma)} T^{\sigma(1)1\pm}(u) \cdots T^{\sigma(\ell)\ell\pm}(q^{-\ell+1} u),
\end{equation}
with $S_\ell$ the symmetric group on $\ell$ objects, to extract the first modes of the quantum determinant and find
\begin{align}\label{eq:qdet}
    \operatorname{qdet}^\pm[0] = t^{\pm N/2}, \qquad \operatorname{qdet}^\pm[1] = q^{\mp 1/2} (1-t^{\mp 1}) H^\pm[1].
\end{align}
It follows that the Bethe subalgebra of the horizontal quantum loop algebra $U_q(\dot{\mathfrak{gl}}_\ell)$ (which includes the quantum determinant) provides a large algebra of commuting Hamiltonians including the defining Hamiltonian $H^-[1]$. By a naive counting, the image of the Bethe subalgebra in the $N$-truncated quotient algebra $U_q^{(N)}(\dot{\mathfrak{gl}}_\ell)$ has $N\ell$ algebraically independent generators, which would provide enough commuting Hamiltonians for Liouville integrability\footnote{We thank J. Lamers and M. Vasilev for this insight.}, since the $K$-theoretic Coulomb branch algebra is generated by the $2N\ell$ algebraically independent generators $Q_i^\alpha$ and $P_i^\alpha$.

\smallskip For the first few generalized Macdonald operators \eqref{eq:macdonald}, one can verify that they commute with the Drinfeld--Jimbo generators \eqref{eq:DJgen1}--\eqref{eq:DJgen2}. We therefore conjecture that the Hamiltonians $H^\pm[n]$ are central in the horizontal quantum loop algebra $U_q(\dot{\mathfrak{gl}}_\ell)$ and hence lie in the horizontal Heisenberg subalgebra defined in \cite{wen:2025}. The theory of \emph{loc.\ cit.}\ would suggest that eigenstates of the Hamiltonians can be given by wreath Macdonald polynomials.

\section{Quantization of physical spin variables}\label{sec:spinVars}

Following the formulas for  the classical case elaborated in \cite{arutyunov:2026}, the spin variables used by Krichever and Zabrodin \cite{krichever:1995} are defined by the expressions
\begin{equation}
    a^\alpha \coloneq L^{0-} \cdots L^{\alpha-1,-} e, \qquad c^\alpha \coloneq u^{\alpha+1,-} \tilde L^{\alpha+1,-} \cdots \tilde L^{\ell-1,-}
\end{equation}
for $\alpha = 0,\dots,\ell-1$, where we use the modified $L$-operator $\tilde L^{\alpha-} \coloneq \mu^\alpha Q^\alpha L^\alpha (Q^{\alpha+1})^{-1}$, which satisfies the equation\footnote{This can be seen as the trigonometric analog of the quantum moment map equations of the chainsaw quiver variety \cite[equation (34)]{finkelberg:2014}.}
\begin{equation}
    q^{1/2} L^{\alpha-} - q^{-1/2} \tilde L^{\alpha-} = e u^{\alpha+1,-}.
\end{equation}
The same telescopic argument as in \cite{arutyunov:2026} yields a similar identity for the total $L$-operators
\begin{equation}\label{eq:Ltot}
        \sum_{\alpha \in \Z/\ell\Z} q^\alpha a^\alpha c^\alpha
        = q^{\ell-1/2} L^{\text{tot},-} - q^{-1/2} \tilde L^{\text{tot},-},
\end{equation}
with $\tilde L^{\text{tot}-} \coloneq \tilde L^{0,-} \cdots \tilde L^{\ell-1,-} = t Q^0 L^{\text{tot}-} (Q^0)^{-1}$. We can solve equation \eqref{eq:Ltot} for the matrix coefficients, obtaining
\begin{equation}
    L_{ij}^{\text{tot}-} = \frac{q^{1/2-\ell}}{1- tq^{1-\ell} Q_i^0/Q_j^0} \sum_{\alpha \in \Z/\ell\Z} q^\alpha a_i^\alpha c_j^\alpha, \qquad \tilde L_{ij}^{\text{tot}-} = \frac{q^{1/2}}{t^{-1} q^{\ell-1} Q_j^0/Q_i^0-1} \sum_{\alpha \in \Z/\ell\Z} q^\alpha a_i^\alpha c_j^\alpha.
\end{equation}

We now comment on the commutation relations of the \emph{physical} spin variables $a^\alpha$ and $c^\alpha$, meaning we restrict ourselves to spin states with index $\alpha \neq 0$. We obtain:
\begin{align}
    Q_i^0 a_j^\alpha &= a_j^\alpha Q_i^0, \label{eq:spinRel1} \\[6pt]
    Q_i^0 c_j^\alpha &= q^{\delta_{ij}} c_j^\alpha Q_i^0, \\[6pt]
    a_a^\alpha a_b^\alpha &= R_{ab}^{00}(q) a_b^\alpha a_a^\alpha, \\[6pt]
    a_a^\alpha a_b^\beta &= q^{1/2} R_{ab}^{00}(q) a_b^\beta a_a^\alpha, & (\alpha < \beta) \\[-2pt]
    c_a^\alpha \bar R_{ba}^{00}(q)^{-1} a_b^\alpha &= q^{-1/2} a_b^\alpha c_a^\alpha - q^{-\alpha} (1-q^{-1}) \bigg[ \Delta_{ab} \tilde L_a^{\text{tot}-} + q^{1/2} \sum_{\rho=0}^{\alpha-1} q^\rho a_b^\rho c_a^\rho \bigg], \\[-3pt]
    c_a^\alpha \bar R_{ba}^{00}(q)^{-1} a_b^\beta &= a_b^\beta c_a^\alpha, & (\alpha < \beta) \\[6pt]
    c_a^\alpha c_b^\alpha &= c_b^\alpha c_a^\alpha \underline R_{ab}^{00}(q)^{-1}, \\[6pt]
    c_a^\alpha c_b^\beta &= q^{-1/2} c_b^\beta c_a^\alpha \underline R_{ab}^{00}(q)^{-1}, & (\alpha < \beta) \label{eq:spinRel2}
\end{align}
with $\Delta = \sum_{i=1}^N e_i^t \otimes e_i$.

\smallskip
The commutation relations \eqref{eq:spinRel1}--\eqref{eq:spinRel2} show how the spin variables commute given definite spin states $\alpha,\beta$. Physically, however, it makes more intuitive sense to ask how the spin variables of two given particles $i,j$ interchange. This question indeed has a nice answer in the case of the physical spin vectors $c_i$, where we observe a simple exchange relation:
\begin{equation}\label{eq:spinExchange}
    c_i^a c_i^b = R^{ab}(-1) c_i^b c_i^a, \qquad c_i^a c_j^b = R^{ab}(Q_j^0/Q_i^0) c_j^b c_i^a, \qquad (i \neq j)
\end{equation}
with $R(u)$ the standard trigonometric $R$-matrix \eqref{eq:trigRMat} for $\mathfrak{gl}_{\ell-1}$. However, a similar formula for the physical spin covectors $a_i$ is not available, repeating the pattern remarked upon after proposition 3.6 of \cite{arutyunov:2025} in the rational degeneration.

\smallskip
We now finish the discussion of the quantization of the physical spin variables by giving their quantum equations of motion (the Heisenberg equations). This can be done simply by utilizing the commutation relations \eqref{eq:oneSiteLmAlg}. For example, we can compute
\begin{align}
    a^\alpha H^-[1] = \operatorname{Tr}_a a_b^\alpha L_a^{\text{tot}-} = q^{1/2} \operatorname{Tr}_a R_{ba}^{00}(q) L_a^{\text{tot}-} \bar R_{ba}^{00}(q)^{-1} a_b^\alpha,
\end{align}
where on the way we have used the identity $\underline R_{ba}^{\beta\alpha}(q)^{-1} e_b = q^{1/2} \bar R_{ab}^{\alpha\beta}(q)^{-1} e_b$. Writing these identities in components, the Heisenberg equations become
\begin{align}
    \dot{Q}_i^0=   [H^-[1],Q_i^0] &= (q^{-1}-1) Q_i^0 L_{ii}^{\text{tot}-}, \label{eq:eoms1} \\[8pt]
    \dot{a}_i^\alpha=  [H^-[1],a_i^\alpha] &= (q^{-1}-1) \sum_{j(\neq i)} L_{ij}^{\text{tot}-} (a_j^\alpha-a_i^\alpha) \frac{1}{1-Q_j^0/Q_i^0}, \\[-2pt]
    \dot{c}_i^\alpha=  [H^-[1],c_i^\alpha] &= (q^{-1}-1) \sum_{j(\neq i)} \bigg( \frac{1}{1-Q_j^0/Q_i^0} c_i^\alpha L_{ij}^{\text{tot}-} - \frac{1}{1-Q_i^0/Q_j^0} c_j^\alpha L_{ji}^{\text{tot}-} \bigg), \label{eq:eoms2}
\end{align}
which give the natural operator ordering for the operator form of the classical equations of motion by Krichever and Zabrodin \cite{krichever:1995}.

\section{Conclusion}

In this paper, we have quantized the trigonometric spin RS model of $N$ particles carrying $\ell$ spin degrees of freedom, starting from its classical description in terms of the $K$-theoretic Coulomb branch of the 4d $\mathcal{N}=2$ necklace quiver gauge theory with $\ell$ nodes of rank $N$ found in \cite{arutyunov:2026}. The main tool of the construction is the algebra of $L$-operators $L^{\alpha\pm}$ of equations \eqref{eq:LpOp} and \eqref{eq:LmOp}. Their commutation relations \eqref{eq:oneSiteLpAlg} and \eqref{eq:oneSiteLmAlg} are governed by a dynamical $R$-matrix $\underline R^{\alpha\beta}(q)$, whose dynamical parameters are the abelianized scalars $Q_i^\alpha$, a constant $R$-matrix $R^{\alpha\beta}(q)$ solving the ordinary Yang--Baxter equation, and the dynamical twist $\bar R^{\alpha\beta}(q)$ relating the two. In this way the necklace quiver is turned into an integrable spin chain of length $\ell$, where the nodes of the quiver play the role of the sites.

\smallskip
We find that the algebra \eqref{eq:totalAlg} of the total $L$-operators $L^{\text{tot}\pm}$ collapses onto the $L$-operator algebra of the spinless model of \cite{arutyunov:2019b}: the twists telescope around the necklace and only the node $\alpha=0$ survives. Consequently, the trace formula \eqref{eq:traceFormula} of \emph{loc.\ cit.} produces commuting Hamiltonians $H^\pm[n]$ for the spin model as well, which we have rewritten through the determinant formula \eqref{eq:determinant} in terms of the generalized Macdonald operators $S^\pm[n]$ of equation \eqref{eq:macdonald}. For $\ell=1$, the operators $S^\pm[n]$ reduce to the Macdonald difference operators.

\smallskip
We have furthermore exhibited the horizontal ($N$-truncated) quantum loop subalgebra
$U^{(N)}_q(\dot{\mathfrak{gl}}_\ell)$ inside the $K$-theoretic Coulomb branch algebra by realizing its Drinfeld--Jimbo generators \eqref{eq:DJgen1}--\eqref{eq:DJgen2} in terms of abelianized monopole operators, and by expressing the low modes of the RTT generator matrices $T^\pm(u)$ via the monopole operators $E^{\alpha,\beta}$ and $F^{\alpha,\beta}$. The quantum Leibniz formula then yields the identification \eqref{eq:qdet} of $H^\pm[1]$ with the first mode of the quantum determinants, so that the defining Hamiltonian $H^-[1]$ of the spin RS model lies in the Bethe subalgebra of the horizontal quantum loop subalgebra. Since the image of this maximal commutative subalgebra in the $N$-truncated quotient naively has $N\ell$ algebraically independent generators, while the quantized abelianized $K$-theoretic Coulomb branch algebra is generated by the $2N\ell$ elements $Q^\alpha_i$, $P^\alpha_i$, this should provide enough commuting Hamiltonians for Liouville integrability. Finally, we have determined the commutation relations \eqref{eq:spinRel1}--\eqref{eq:spinRel2} of the physical spin variables, which form a quadratic algebra controlled by the $R$-matrices of section \ref{sec:LOpAlgebra}, and we have written down the quantum equations of motion \eqref{eq:eoms1}--\eqref{eq:eoms2}, which supply the correct operator ordering for the Krichever--Zabrodin equations of motion \cite{krichever:1995}.

\smallskip
Several questions are left open by our analysis. The most immediate one is our conjecture that all Hamiltonians $H^\pm[n]$, and not only $H^\pm[1]$, are central in the horizontal quantum loop algebra, which we have verified for the first few $n$ by direct computation. A proof would place the whole family inside the horizontal Heisenberg subalgebra of \cite{wen:2025} and would suggest wreath Macdonald polynomials as eigenstates of the Hamiltonians $H^\pm[n]$, thereby solving the spectral problem of the trigonometric spin RS model in closed form. The $\ell=1$ specialization of this statement is the well-known fact that Macdonald polynomials diagonalize the Macdonald operators.

\smallskip
As suggested by D. Gaiotto\footnote{Private communication.}, one expects the interpretation of the $L$-operators in terms of the 4d $\mathcal{N}=2$ necklace quiver gauge theory to be that they are line interfaces between certain Gukov--Witten surface defects. Mathematically, this expectation should be formulated as the statement that the $L$-operators are morphisms in the $K$-theoretic version of Webster's extended BFN category constructed in \cite[\S 3.2, \S 3.3]{webster:2024}, whose objects would be Gukov--Witten surface defects and whose morphisms would be line interfaces. This expectation holds because Gukov--Witten surface defects wrapping $S_\theta^1$ in 4d should go to vortex line defects in 3d after sending the radius of $S_\theta^1$ to zero, and the objects of the (homological) extended BFN category were explicitly identified with vortex line defects in \cite{webster:2024b}. Future work thus includes realizing the construction of the $L$-operators in the $K$-theoretic version of Webster's extended BFN category \cite{webster:2024}.

\smallskip
The treatment of hermitian representations is also left open for future analysis. We expect that an application of the framework of Schur quantization \cite{gaiotto:2024b} will be fruitful in this regard. Another relevant direction in the treatment of hermitian representations is to work out the connection to the trigonometric spin RS model defined in \cite{lamers:2022}. Finally, we remark that the investigation of the elliptic spin RS model remains open, because few descriptions of elliptic Coulomb branches \cite{finkelberg:2020} have been explored sufficiently.

\appendix

\acknowledgments

We thank Davide Gaiotto, Jules Lamers, Elli Pomoni, Alexander Shapiro, Jörg Teschner, Mikhail Vasilev, Ben Webster, and Yegor Zenkevich for fruitful discussions. G.A. acknowledges support by the DFG under Germany's Excellence Strategy -- EXC 2121 ``Quantum Universe'' -- 390833306. G.A. and L.H. acknowledge support by the DFG -- SFB 1624 -- ``Higher structures, moduli spaces and integrability'' -- 506632645. R.K.’s research is partially funded by the Deutsche Forschungsgemeinschaft (DFG, German
Research Foundation) -- Projektnummer 417533893/GRK2575 “Rethinking Quantum Field Theory”. This research was supported in part by Perimeter Institute for Theoretical Physics. Research at Perimeter Institute is supported by the Government of Canada through the Department of Innovation, Science and Economic Development and by the Province of Ontario through the Ministry of Research, Innovation and Science.




\bibliographystyle{JHEP}
\bibliography{bibliography.bib}

\end{document}